\documentclass[12pt,a4paper]{article}
\usepackage[cp1251]{inputenc}
\usepackage[T2A]{fontenc}
\usepackage{amsfonts}
\usepackage{amsmath}
\usepackage{amssymb}
\usepackage[russian,english]{babel}
\usepackage{graphicx}
\newtheorem{law1}{Corollary}
\newtheorem{law2}{Theorem}
\begin{document}
\title{A sequential growth dynamics for a directed acyclic dyadic graph} 

\author{Alexey L. Krugly\thanks{Department of applied mathematics and computer science, Scientific Research Institute for System Analysis of the Russian Academy of Science, 117218, Nahimovskiy pr., 36, k. 1, Moscow, Russia; akrugly@mail.ru.}}
\date{} \maketitle
\begin{abstract}
A model of discrete spacetime on a microscopic level is considered. It is a directed acyclic dyadic graph (an x-graph). The dyadic graph means that each vertex possesses no more than two incident incoming edges and two incident outgoing edges. This model is the particular case of a causal set because the set of vertices of x-graph is a causal set. The sequential growth dynamics is considered. This dynamics is a stochastic sequential additions of new vertices one by one. A new vertex can be connected with existed vertex by an edge only if the existed vertex possesses less than four incident edges. There are four types of such additions. The probabilities of different variants of addition of a new vertex depend on the structure of existed x-graph. These probabilities are the functions of the probabilities of random choice of directed paths in the x-graph. The random choice of directed paths is based on the binary alternatives. In each vertex of the directed path we choose one of two possible edges to continue this path. It is proved that such algorithm of the growth is a consequence of a causality principle and some conditions of symmetry and normalization. The probabilities are represented in a matrix form. The iterative procedure to calculate probabilities is considered. Elementary evolution operators is introduced. The second variant to calculate probabilities is based on these elementary evolution operators.


\end{abstract}

\section{Introduction}
\label{intro}
\begin{figure}[ht]
	\centering	
		\includegraphics[trim=8cm 18cm 8cm 4cm]{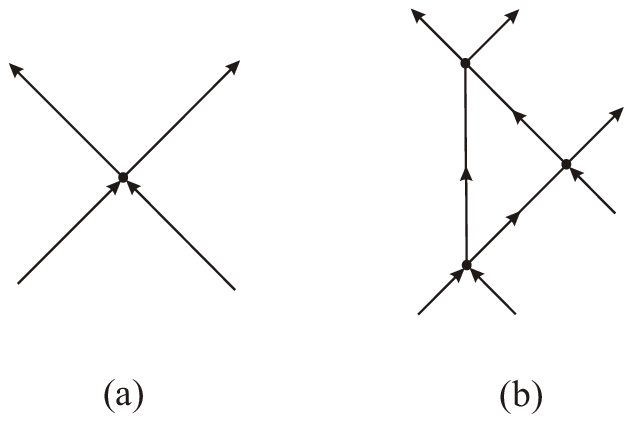}
	\caption{(a) An x-structure. (b) The x-graph with 3 vertices.}
	\label{fig:fig1}
\end{figure}

By assumption spacetime is discrete on a microscopic level. Consider a particular model of such discrete pregeometry. This is a directed dyadic acyclic graph. All edges are directed. The dyadic graph means that each vertex possesses two incident incoming edges and two incident outgoing edges. A vertex with incident edges forms an x-structure (Fig.\ \ref{fig:fig1}a).
This model was suggested by D. Finkelstein in 1988 \cite{Fink88}. The acyclic graph means that there is not a directed loop. This graph is called an x-graph. Consider an example of x-graph with 3 vertices (Fig.\ \ref{fig:fig1}b). There is one loop. But this is not a directed loop. This loop includes 2 edges in the same direction and 1 edge in opposite direction.

This model is the particular case of a causal set. A causal set is a pair ($\mathcal{C}$, $\prec$), where $\mathcal{C}$ is a set and $\prec$ is a binary relation on $\mathcal{C}$ satisfying the following properties ($x,\ y,\ z$ are general elements of $\mathcal{C}$):
\begin{equation}
\label{eq:I1.1} x\nprec x\qquad \textrm{(irreflexivity),}
\end{equation}
\begin{equation}
\label{eq:I1.2} \{x\mid(x\prec y)\wedge(y\prec x)\}=\emptyset \qquad \textrm{(acyclicity),}
\end{equation}
\begin{equation}
\label{eq:I1.3} (x\prec y)\wedge(y\prec z)\Rightarrow(x\prec z)\qquad \textrm{(transitivity),}
\end{equation}
\begin{equation}
\label{eq:I1.4} \mid\mathcal{A}(x, y)\mid<\infty\qquad\textrm{(local finiteness),}
\end{equation}
$$
\textrm{where\ }\mathcal{A}(x,\ y)=\{z\mid x\prec z \prec y\}\textrm{.}
$$
The first three properties are irreflexivity,  acyclicity, and transitivity. These are the same as for events in Minkowski spacetime. $\mathcal{A}(x,\ y)$ is called an Alexandrov set of the elements $x$ and $y$ or a causal interval or an order interval. In Minkowski spacetime, an Alexandrov set of any pair of events is an empty set or a set of continuum. The local finiteness means that an Alexandrov set of any elements is finite. The physical meaning of this binary relation $\prec$ is causal or chronological order. By assumption a causal set describes spacetime and matter on a microscopic level. In the x-graph, the set of vertexes and the set of edges are causal sets. 

A causal set approach to quantum gravity has been introduced by G. 't Hooft \cite{tHooft78} and J. Myrheim \cite{Myrheim78} in 1978. The term `causal set' was proposed in 1987 \cite{BMSorkin}. There are reviews of a causal set program \cite{Sorkin2005,Dowker2006,Henson2009,Wallden2010}. In most of papers, the connection of the causal set and continuous spacetime is considered. The aim is to deduce continuous spacetime and its properties (for example, the dimensionality $3+1$) as some approximation of the causal set (see e.g. \cite{BG2010}) or to consider some particular problem (see e.g. \cite{0302009}). However, if we consider the causal set as a description of a most deep level of the universe, the causal set must describe matter. In some papers, quantum fields on a background causal set are considered (see e.g. \cite{0806.3083,0801.0240,0905.1506,1107.0698}). This approach can be fruitful as approximation. But, in the self-consistent causal set theory, the matter must be a property of the causal set. 

In quantum field theory, the properties of particles are considered as manifestations of symmetry. In the considered model, particles can be repetitive symmetrical structures of the x-graph. The symmetry is defined for an infinite perfect x-graph \cite{Krugly2010}. Similarly, in the crystallography, the symmetry is defined for infinite perfect crystals. Number the set of edges $\{e_i\}$ of the x-graph. We can use any numbering of the edges if the different edges have the different numbers. This is an admissible numbering. The renumbering of the edges is a map $\mathbf{F}(a)=b$, when $a$ is the old number of the edge and $b$ is a new number if old and new numberings are admissible. The sets $\{ a\}$ and $\{ b\}$ of old and new numbers may be different. For example, $\{ a\}$ is a set of integers and $\{ b\}$ is a set of odd integers. If these sets coincide, $\mathbf{F}(a)=b$ is a permutation. We can consider the symmetry of the x-graph as a property of the partial order of its edges. Consider the permutation $\mathbf{F}$ of the numbers of the edges in the x-graph such that $e_{\mathbf{F}(a)}\prec e_{\mathbf{F}(b)}$ if and only if $e_a\prec e_b$. Then the permutation $\mathbf{F}$ is called a symmetry of this x-graph. We can consider the order reversibility (the change of directions of all edges). This is the discrete analog of the time reversibility. In \cite[section~4.5]{Krugly2010}, there are the classification of groups of symmetries and their properties for the considered model. In \cite[section~4.6]{Krugly2010}, there are some examples of infinite repetitive symmetrical x-graphs. Real structures are finite. They can have an approximate symmetry. Such structures must emerge as a consequence of dynamics.

The goal of this model is to describe particles as some repetitive symmetrical self-organized structures of the x-graph. This self-organization must be the consequence of dynamics. In this paper, I introduce an example of dynamics.

\section{The sequential growth}
\label{SG}
The model of the universe is an infinite x-graph. Each directed path can be infinitely continued in both directions in the x-graph. But any observer can only actually know a finite number of facts. Then an observer can only know a finite x-graph. In a graph theory, by definition, an edge is a relation of two vertices. Consequently some vertices of a finite x-graph have less than four incident edges. These vertices have free valences instead the absent edges. These free valences are called external edges as external lines in Feynman diagrams. They are figured as edges that are incident to only one vertex. There are two types of external edges: incoming external edges and outgoing external edges. The x-graph with 3 vertices (Fig.\ \ref{fig:fig1}b) possesses 3 incoming external edges and 3 outgoing external edges. We can prove that the number of incoming external edges is equal to the number of outgoing external edges \cite{1008.5169}\footnote{It should be noted that a set of halves of edges is considered in papers \cite{1008.5169,1004.5077,1106.6269}. The halves of edges as basic objects were introduced by D. Finkelstein and G. McCollum in 1975 \cite{FinkMcC1}. By some reasons, it is convenient to break the edge into two halves of which the edge is regarded as composed. The set of halves of edges in papers \cite{1008.5169,1004.5077,1106.6269} is isomorphic to the considered x-graph.}.

Each x-graph is a model of a part of some process. The task is to predict the future stages of this process or to reconstruct the past stages. We can reconstruct the x-graph step by step. The minimal part is a vertex. We start from some given x-graph and add new vertices one by one. This procedure is proposed in papers of author \cite{Krugly1998,Krugly2002}. Similar procedure and the term `a classical sequential growth dynamics' is proposed by D. P. Rideout and R. D. Sorkin \cite{RideoutSorkin} for other model of causal sets.

We can add a new vertex to external edges. This procedure is called an elementary extension. There are four types of elementary extensions \cite{1004.5077}. There are two types of elementary extensions to outgoing external edges (Fig.\ \ref{fig:fig2}a and\ \ref{fig:fig2}b). This is a reconstruction of the future of the process.  In this and following figures the x-graph $\mathcal{G}$ is represented by a rectangle because it can have an arbitrary structure. The edges that take part in the elementary extension are figured by bold arrows. First type is an elementary extension to two outgoing external edges (Fig.\ \ref{fig:fig2}a).
\begin{figure}[t]
	\centering
\includegraphics[width=3.5cm,trim=8cm 12cm 8cm 2cm]{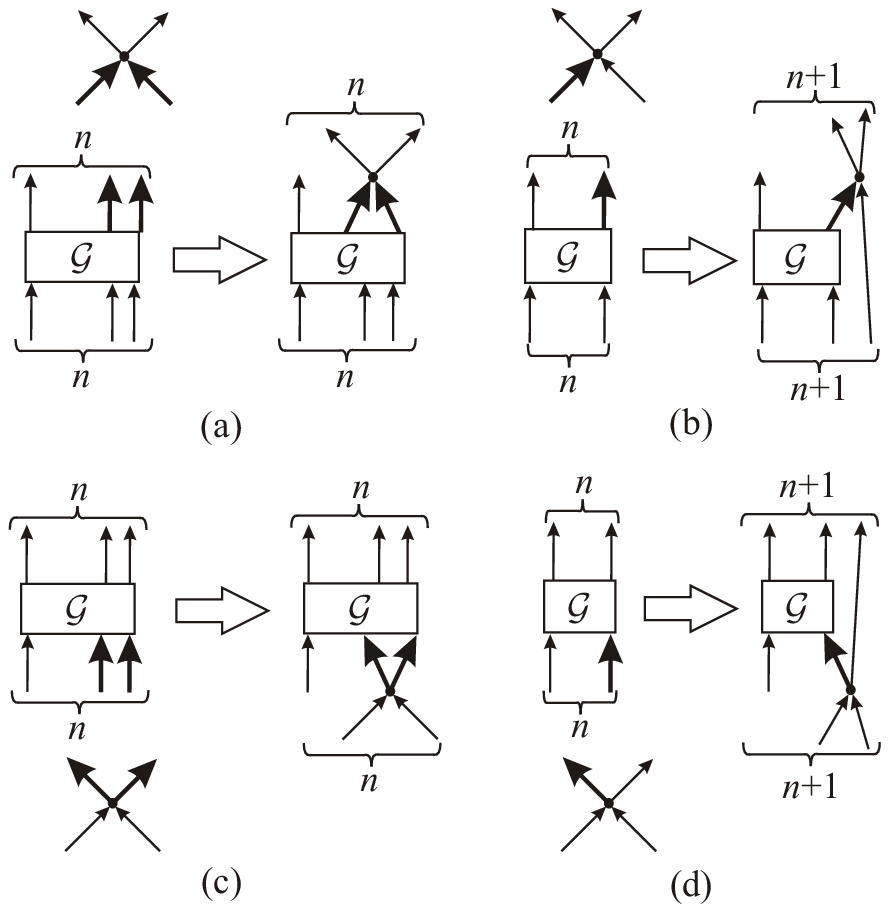}
\caption{\label{fig:fig2}The types of elementary extensions: (a) the first type, (b) the second type, (c) the third type, and (d) the fourth type.}
\end{figure}
Second type is an elementary extension to one outgoing external edge (Fig.\ \ref{fig:fig2}b). Similarly, there are two types of elementary extensions to incoming external edges (Fig.\ \ref{fig:fig2}c and\ \ref{fig:fig2}d). These elementary extensions reconstruct the past evolution of the process. Third type is an elementary extension to two incoming external edges (Fig.\ \ref{fig:fig2}c). Fourth type is an elementary extension to one incoming external edge (Fig.\ \ref{fig:fig2}d). In the elementary extensions of the first or third types, the number $n$ of incoming or outgoing external edges is not changed. These elementary extensions describe the interior evolution of the process. In the elementary extensions of the second or fourth types, the numbers $n$ of incoming external edges and outgoing external edges have increased by 1. These elementary extensions describe the interactions of the process and environment. If we consider the x-graph as a partially ordered set of vertices, the elementary extension of the first or second types are the addition of a maximal vertex, and the elementary extension of the third or fourth types are the addition of a minimal vertex. We can prove that we can get every connected x-graph by a sequence of elementary extensions of these four types \cite[Teorem~2]{1008.5169}.

Consider an interpretation of the sequential growth. There are two concepts of time. In the first concept, the future does not exist and emerges from the present. In the second concept, the past and the future exist, are determined, and are changeless. For example, these two concepts are described in the introduction of \cite{HR1956}. The first concept of the future for a sequential growth is introduced in \cite{0703098}. In this paper, I consider the second concept. This means that the unique infinite x-graph of the universe exists. We have the following assumption. Any finite subgraph of the x-graph of the universe has the certain structure, and we can determine this structure. Consequently, the structure of any finite subgraph of the x-graph of the universe is an observable. An observer observes this structure by the sequential growth. The addition of a new vertex is not an appearance of a new part of the infinite x-graph of the universe. This is an appearance of new information about the existing infinite x-graph of the universe. The sequential growth is a growth of information about the existing universe.

There are two times in this model. The first time is the partial order of vertices and edges of the x-graph. This is the symmetrical reversible time of an object. The second time is the linear order of the elementary extensions during the sequential growth. This is the asymmetrical irreversible time of an observer. The direction of the time arrow of an observer is the direction of the growth of information.

The minimal part of information is one vertex. By assumption, observer can randomly initiate one elementary extension and can determine the exact result of this elementary extension. This procedure is called an elementary measurement. In general case, observer cannot forecast the exact result of the elementary measurement. He can only calculate probabilities of different variants. Otherwise, he can calculate the exact structure of the whole universe. In quantum theory, a set of results of sequential measurements is a classical stochastic sequence. Similarly, a sequence of the elementary extensions is a classical stochastic sequence.

The aim of the dynamics is to calculate the probabilities of the elementary extensions.

\section{An algorithm to calculate the probabilities}
\label{ACP}
\begin{figure}[t]
	\centering	
		\includegraphics[width=4cm,trim=8cm 14cm 8cm 4cm]{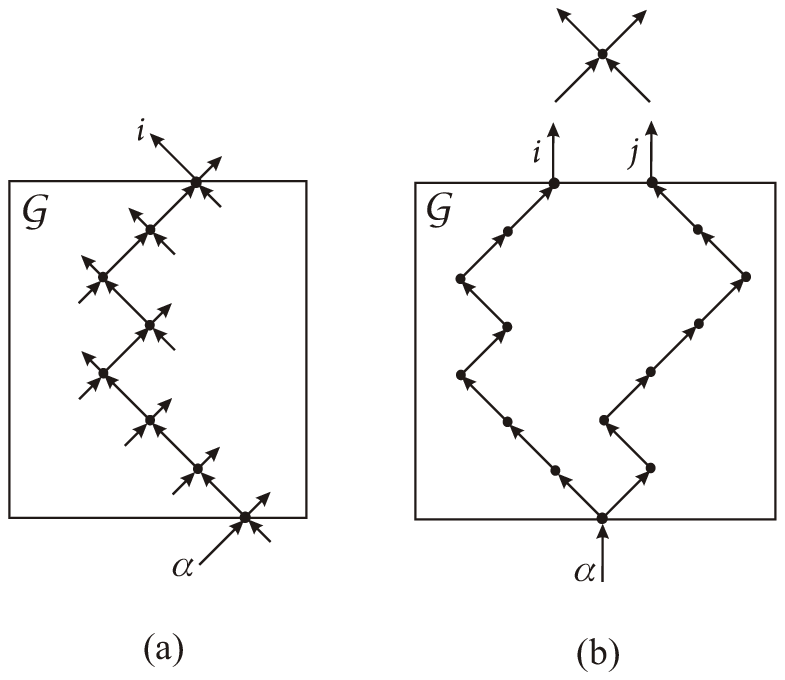}
	\caption{(a) A choice of a directed path is a sequence of binary alternatives. (b) A new loop is generated by a new vertex.}
	\label{fig:fig3}
\end{figure}

Consider a directed path. Number outgoing external edges by Latin indices. Number incoming external edges by Greek indices. Latin and Greek indices range from 1 to $n$, where $n$ is the number of outgoing or incoming external edges. If we choose a directed path from any incoming external edge number $\alpha$, we must choose one of two edges in each vertex (Fig.\ \ref{fig:fig3}a). Assume the equal probabilities for both outcomes independently on the structure of the x-graph. Then this probability is equal to $1/2$. Consequently if a directed path includes $k$ vertices, the choice of this path has the probability $2^{-k}$. We have the same choice for opposite directed path.

Introduce an amplitude $a_{i\alpha }$ of causal connection of the outgoing external edge number $i$ and the incoming external edge number $\alpha$. By definition, put
\begin{equation}
\label{eq:1} a_{i\alpha}= a_{\alpha i}=\sum_{m=1}^M 2^{-k(m)}\textrm{,}
\end{equation}
where $M$ is the number of directed paths from the incoming external edge number $\alpha $ to the outgoing external edge number $i$, and $k(m)$ is the number of vertices in the path number $m$. This definition has clear physical meaning. The causal connection of two edges is stronger if there are more directed paths between these edges and these paths are shorter. Throughout the paper these amplitudes are called amplitudes for simplicity.

Consider a following algorithm to calculate the probabilities of elementary extensions \cite{1106.6269}. Define the algorithm using the amplitudes. There are three steps.

The first step is the choice of the elementary extension to the future or to the past. By definition, the probability of this choice is $1/2$ for both outcomes.

A new vertex is added to one or two external edges. The second step is the equiprobable choice of one external edge that takes part in the elementary extension. This is an outgoing external edge if we have chosen the future evolution in the first step. Otherwise this is an incoming external edge. The probability of this choice is $1/n$ for each outcome.

The third step is the choice of second external edge. Denote by $p_{ij}$ the probability to choose the outgoing external edge number $j$ if we have chosen the outgoing external edge number $i$ in the second step. By definition, put 
\begin{equation}
\label{eq:2} p_{ij}=\sum_{\alpha=1}^n a_{i\alpha} a_{\alpha j}\textrm{.}
\end{equation}

Consider the meaning of this definition. The addition of a new vertex to two external edges forms a set of loops (Fig.\ \ref{fig:fig3}b). Each loop is formed by two directed paths. We can describe a loop by a weight that is a product of probabilities of these paths. The probability of the elementary extension is directly proportional to the sum of weights of new loops that are generated by this elementary extension.

Similarly,
\begin{equation}
\label{eq:3} p_{\alpha \beta}=\sum_{i=1}^n a_{\alpha i} a_{i\beta}\textrm{,}
\end{equation}
where $p_{\alpha \beta}$ is the probability to add a new vertex to two incoming external edges numbers $\alpha$ and $\beta$.

The sum of probabilities of all directed paths from any edge is equal to 1. We get the right normalization if we put the following definition for the probability to add a new vertex to one outgoing or incoming external edge number $i$ or $\alpha$, respectively.
\begin{equation}
\label{eq:4} p_{ii}=\sum_{\alpha=1}^n a_{i\alpha} a_{\alpha i}\textrm{,}
\end{equation}
\begin{equation}
\label{eq:5} p_{\alpha \alpha}=\sum_{i=1}^n a_{\alpha i} a_{i\alpha }\textrm{.}
\end{equation}

We can express these equations in a matrix form. Introduce a matrix $\mathbf{a}$ of amplitudes. All matrixes are denoted by bold Latin letters. An element $a_{i\alpha}$ of this matrix is equal to the amplitude of causal connection of the outgoing external edge number $i$ and the incoming external edge number $\alpha$. The matrix $\mathbf{a}$ is a square matrix of size $n$. Introduce a matrix $\mathbf{p}_f$ of probabilities of elementary extensions to the future and a matrix $\mathbf{p}_p$ of probabilities of elementary extensions to the past. An element number $ij$ of $\mathbf{p}_f$ is equal to $p_{ij}$. An element number $\alpha \beta$ of $\mathbf{p}_p$ is equal to $p_{\alpha \beta}$. We have the matrix form of (\ref{eq:2}) and (\ref{eq:4})
\begin{equation}
\label{eq:6} \mathbf{p}_f= \mathbf{a}\mathbf{a}^T \textrm{,}
\end{equation}
and the matrix form of (\ref{eq:3}) and (\ref{eq:5})
\begin{equation}
\label{eq:7} \mathbf{p}_p= \mathbf{a}^T \mathbf{a} \textrm{.}
\end{equation}
The sum of the elements in each row and in each column is equal to 1 for the matrixes $\mathbf{a}$, $\mathbf{p}_f$, and $\mathbf{p}_p$.

\section{Physical foundations of $p_{ij}$}
\label{PF}
The physical foundations of the first and second steps of the algorithm are trivial. The introduced algorithm to calculate $p_{ij}$ is based on the next physical assumptions.
\begin{itemize}
\item Causality.
\item Symmetry.
\item Normalization.
\end{itemize}
The symmetry and the normalization are trivial.

In this model, causality is defined as the order of vertices and edges. But the causality has a real physical meaning only if the dynamics agrees with causality. The probability to add a new vertex to the future can only depend on the subgraph that precedes this vertex \cite{RideoutSorkin}. Similarly, the probability to add a new vertex to the past can only depend on the subgraph that follows this vertex. This is the causality principle for the considered model.

Consider the x-graph $\mathcal{G}$. By definition, put $\mathcal{P}(v)=\{ v_i\in \mathcal{G}\mid v_i\prec v\}$. The set $\mathcal{P}(v)$ is called the past set of the vertex $v$. By definition, put $\mathcal{F}(v)=\{ v_i\in \mathcal{G}\mid v\prec v_i\}$. The set $\mathcal{F}(v)$ is called the future set of the vertex $v$.

\begin{law2}\label{T1}
Consider the x-graph $\mathcal{G}$ that consists of the set $\{ v\}$ of vertexes. The cardinality $|\{ v\}|=N$. Consider the conditional probability $p_{ij}$ to add a new maximal vertex $v_{N+1}$ to the outgoing external edges numbers $i$ and $j$ if we choose the outgoing external edge number $i$. The edges $i$ and $j$ can coincide. If $p_{ij}$ is a function of $\mathcal{P}(v_{N+1})$ (causality), $p_{ij}=p_{ji}$ (symmetry), and the normalization constant is $n^{-1}$ (normalization), then $p_{ij}=\sum_{\alpha=1}^n a_{i\alpha} a_{\alpha j}$. 
\end{law2}

\textit{Proof.} The proof is by induction on $N$.

Consider an x-structure (Fig.\ \ref{fig:fig1}a). Number the outgoing external edges by 1 and 2. We have $p_{11}=p_{12}=p_{22}$ by the symmetry and causality. We have $p_{11}+p_{12}=1$ by the normalization. We get $p_{11}=p_{12}=p_{22}=1/2$.

\begin{figure}[t]
	\centering	
		\includegraphics[width=4cm,trim=8cm 14cm 8cm 5cm]{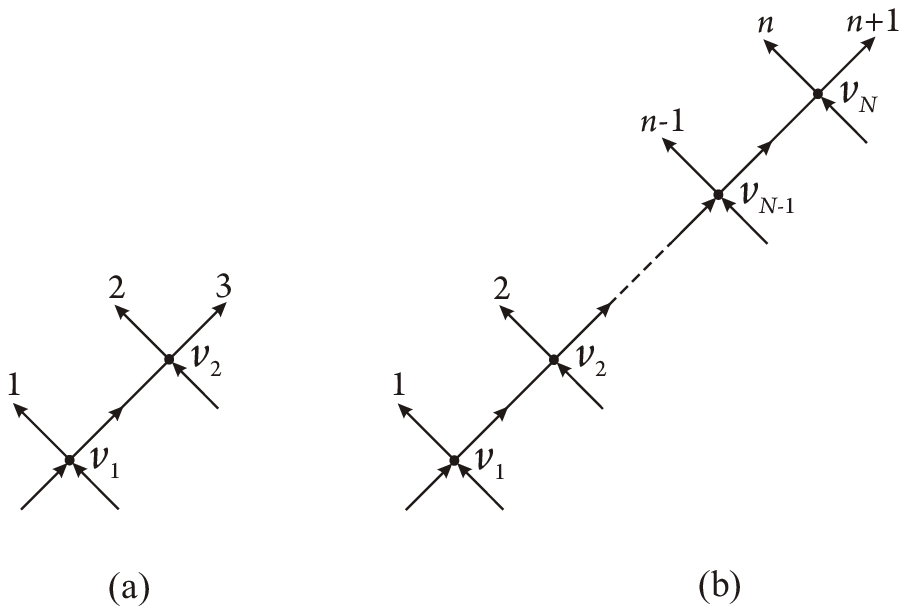}
	\caption{(a) The tree with two vertexes. (b) The tree with $N$ vertexes.}
	\label{fig:fig4}
\end{figure}

Consider the tree $\mathcal{T}_2$ with two vertices (Fig.\ \ref{fig:fig4}a).
We can get this tree by addition of a maximal vertex $v_2$ to the x-structure. Consider the addition of a third vertex $v_3$ to the outgoing external edge number 1. In this case, the past set of $v_3$ does not include $v_2$. Consequently the probability $p_{11}$ of this elementary extension does not depend on $v_2$ by the causality principle. We get $p_{11}=1/2$. We have $p_{12}=p_{13}$ and $p_{22}=p_{23}=p_{33}$ by the symmetry. We have $p_{11}+p_{12}+p_{13}=1$ by the normalization. We get $p_{12}=p_{13}=1/4$. We have $p_{12}+p_{22}+p_{23}=1$ by the normalization. We get $p_{22}=p_{23}=p_{33}=3/8$.

Consider the tree $\mathcal{T}_{N}$ with $N$ vertexes (Fig.\ \ref{fig:fig4}b). We can get $\mathcal{T}_N$ by an addition of a maximal vertex $v_N$ to the tree $\mathcal{T}_{N-1}$ that consists of $N-1$ vertices. Denote by $n$ the cardinality of the set of outgoing external edges for $\mathcal{T}_{N-1}$. Number these outgoing external edges from 1 to $n$ such that $v_N$ is added to the edge number $n$. Number the new outgoing external edges of $\mathcal{T}_N$ by $n$ and $n+1$. Consider the addition of a new maximal vertex $v_{N+1}$ to the outgoing external edges numbers $i<n$ and $j<n$. In this case, the past set of $v_{N+1}$ does not include $v_N$. Consequently $p_{ij}\ (i<n,\ j<n)$ for $\mathcal{T}_N$ and $\mathcal{T}_{N-1}$ are the same by the causality principle. We have $2(n+1)$ new unknown probabilities and $n+1$ normalization conditions. But only $n+1$ unknown probabilities are different by the symmetry. Using the normalization and the symmetry of the new outgoing external edges, we get
\begin{equation}
\label{eq:PF1} p_{in}=p_{i(n+1)}=1/2(1-\sum_{j=1}^{n-1}p_{ij})\textrm{,}
\end{equation}
where $i<n$.
Using equation (\ref{eq:PF1}), the normalization, and the symmetry of the new outgoing external edges, we get
\begin{equation}
\label{eq:PF2} p_{nn}=p_{n(n+1)}=p_{(n+1)(n+1)}=1/2(1-\sum_{i=1}^{n-1}p_{in})\textrm{.}
\end{equation}
This unique solution satisfies the causality principle, the normalization, and the symmetry. Equation (\ref{eq:2}) also satisfies the causality principle, the normalization, and the symmetry. Consequently this solution coincides with equation (\ref{eq:2}).  

We consider the tree for simplicity, and do not use the structure of the x-graph. Consider a general case.

By the inductive assumption, the theorem is truth for any x-graph $\mathcal{G}_{N-1}$ that consists of $N-1$ vertices. Consider any x-graph $\mathcal{G}_N$ that consists of $N$ vertices. We can get this x-graph by an addition a new vertex $v_N$ to some $\mathcal{G}_{N-1}$. Let $v_N$ be a maximal vertex. If $v_N$ is not a maximal vertex, choose some maximal vertex $\tilde{v}_N$ in $\mathcal{G}_N$ and remove $\tilde{v}_N$. We get $\mathcal{\tilde{G}}_{N-1}$. It can be unconnected. The theorem is truth for $\mathcal{\tilde{G}}_{N-1}$ by assumption. Add $\tilde{v}_N$ to $\mathcal{\tilde{G}}_{N-1}$. There are two cases. In the first case, $\tilde{v}_N$ is added to two outgoing external edges as for an elementary extension of the first type (Fig.\ \ref{fig:fig2}a). In the second case, $\tilde{v}_N$ is added to one outgoing external edge as for an elementary extension of the second type (Fig.\ \ref{fig:fig2}b). Denote by $n$ the cardinality of the set of outgoing external edges for $\mathcal{\tilde{G}}_{N-1}$.

In the first case, number these outgoing external edges from 1 to $n$ such that $\tilde{v}_N$ is added to the edges numbers $n-1$ and $n$. Number the new outgoing external edges of $\mathcal{G}_N$ by $n-1$ and $n$. The probabilities $p_{ij}\ (i<n-1,\ j<n-1)$ for $\mathcal{G}_N$ and $\mathcal{\tilde{G}}_{N-1}$ are the same by the causality principle. Using the normalization and the symmetry of the new outgoing external edges, we get  
\begin{equation}
\label{eq:PF3} p_{in}=p_{i(n-1)}=1/2(1-\sum_{j=1}^{n-2}p_{ij})\textrm{,}
\end{equation}
where $i<n-1$.
Using equation (\ref{eq:PF3}), the normalization, and the symmetry of the new outgoing external edges, we get
\begin{equation}
\label{eq:PF4} p_{(n-1)(n-1)}=p_{(n-1)n}=p_{nn}=1/2(1-\sum_{i=1}^{n-2}p_{in})\textrm{.}
\end{equation}

In the second case, number the outgoing external edges of $\mathcal{\tilde{G}}_{N-1}$ from 1 to $n$ such that $\tilde{v}_N$ is added to the edge number $n$. Number the new outgoing external edges of $\mathcal{G}_N$ by $n$ and $n+1$. The probabilities $p_{ij}\ (i<n,\ j<n)$ for $\mathcal{G}_N$ and $\mathcal{\tilde{G}}_{N-1}$ are the same by the causality principle. Using the normalization and the symmetry of the new outgoing external edges, we get equations (\ref{eq:PF1}) - (\ref{eq:PF2}). $\Box$

\begin{law1}\label{C1}
Consider the conditional probability $p_{\alpha \beta}$ to add a new minimal vertex $v_{N+1}$ to the incoming external edges numbers $\alpha$ and $\beta$ if we choose the incoming external edge number $\alpha$. The edges $\alpha$ and $\beta$ can coincide. If $p_{\alpha \beta}$ is a function of $\mathcal{F}(v_{N+1})$ (causality), $p_{\alpha \beta}=p_{\beta \alpha}$ (symmetry), and the normalization constant is $n^{-1}$ (norvalization), then $p_{\alpha \beta}=\sum_{i=1}^n a_{\alpha i} a_{i\beta}$. 
\end{law1}

The proof is the same.

\section{An algorithm to calculate the matrix of amplitudes}
\label{MA}
We can calculate the probability of any elementary extension if we can calculate the matrix of amplitudes for every connected x-graph. Consider an iterative procedure for this matrix. This procedure starts from the x-graph that consists of 1 vertex. This is the x-structure (Fig.\ \ref{fig:fig1}a). We have for its matrix of amplitudes
\begin{equation}
\label{eq:8}
\begin{array}{cccc}
\mathbf{a}(1) &=&\left( \begin{array}{cc} 1/2& 1/2 \\ 1/2 & 1/2 \end{array} \right) &\textrm{.}
\end{array}
\end{equation}

By \cite[Teorem~2]{1008.5169} we can get every connected x-graph from the x-structure by a sequence of elementary extensions of the considered four types. Consider the transformations of the matrix of amplitudes for each type of elementary extension. Consider the x-graph that consists of $N$ vertices. Denote by $n$ the number of outgoing or incoming external edges. We get the x-graph that consists of $N+1$ vertices by any elementary extension.

First type is an elementary extension to the future (Fig.\ \ref{fig:fig2}a). Two outgoing external edges numbers $i$ and $j$ become internal edges. We get two free numbers of outgoing external edges: $i$ and $j$. Two new outgoing external edges appear. Number these new outgoing external edges by $i$ and $j$. New outgoing external edges are included in the same paths. These paths are all paths in which the old outgoing external edges numbers $i$ and $j$ are included. These paths pass through one new vertex. Then we must multiply by $1/2$. We get for the elements of rows numbers $i$ and $j$ of $\mathbf{a}(N+1)$ 
\begin{equation}
\label{eq:9}
a_{i\alpha}(N+1)=a_{j\alpha}(N+1)=1/2(a_{i\alpha}(N)+ a_{j\alpha}(N))\textrm{,}
\end{equation}
where $i$ and $j$ are fixed, and $\alpha$ ranges from $1$ to $n$. Other rows and the size of matrix of amplitudes are not changed.

Second type is an elementary extension to the future too (Fig.\ \ref{fig:fig2}b). One outgoing external edge number $i$ becomes an internal edge. We get $i$ as free number of an outgoing external edge. Two new outgoing external edges and one new incoming external edge appear. Number these new outgoing external edges by $i$ and $n+1$, and new incoming external edge by $n+1$. New outgoing external edge number $i$ is included in the same paths as the old outgoing external edge number $i$. These paths pass through one new vertex. Then we must multiply by $1/2$. We get for the elements of the row number $i$ of $\mathbf{a}(N+1)$  
\begin{equation}
\label{eq:10}
a_{i\alpha}(N+1)=(1/2)a_{i\alpha}(N) \textrm{,}
\end{equation}
where $i$ is fixed, and $\alpha$ ranges from $1$ to $n$. New outgoing external edge number $n+1$ is included in the same paths as new outgoing external edge number $i$. We get new row number $n+1$ with the following elements.
\begin{equation}
\label{eq:11}
a_{(n+1) \alpha}(N+1)=a_{i\alpha}(N+1) \textrm{,}
\end{equation}
where $i$ is fixed, and $\alpha$ ranges from $1$ to $n$. The new incoming external edge number $n+1$ is connected by directed paths only with the outgoing external edges numbers $i$ and $n+1$. Each connection includes one path that passes through one vertex. We get new column number $n+1$ with the following elements.
\begin{equation}
\label{eq:15}
a_{i(n+1)}(N+1)= a_{(n+1) (n+1)}(N+1)=1/2 \textrm{,}
\end{equation}
where $i$ is fixed.
\begin{equation}
\label{eq:16}
a_{r(n+1)}(N+1)=0 \textrm{,}
\end{equation}
where $r$ ranges from 1 to $i-1$ and from $i+1$ to $n$. The size of matrix of amplitudes is increased by 1 from $n$ to $n+1$.

Third type is an elementary extension to the past (Fig.\ \ref{fig:fig2}c). Two incoming external edges numbers $\alpha$ and $\beta$ become internal edges. We get two free numbers of incoming external edges: $\alpha$ and $\beta$. Two new incoming external edges appear. Number these new incoming external edges by $\alpha$ and $\beta$. New incoming external edges are included in the same paths. These paths are all paths in which the old incoming external edges numbers $\alpha$ and $\beta$ are included. These paths pass through one new vertex. Then we must multiply by $1/2$. We get for the elements of column numbers $\alpha$ and $\beta$ of $\mathbf{a}(N+1)$ 
\begin{equation}
\label{eq:17}
a_{r\alpha}(N+1)=a_{r\beta}(N+1)=1/2(a_{r\alpha}(N)+ a_{r\beta}(N))\textrm{,}
\end{equation}
where $\alpha$ and $\beta$ are fixed, and $r$ ranges from 1 to $n$. Other columns and the size of matrix of amplitudes are not changed.

Fourth type is an elementary extension to the past too (Fig.\ \ref{fig:fig2}d). One incoming external edge number $\alpha$ becomes an internal edge. We get $\alpha$ as free number of incoming external edges. Two new incoming external edges and one new outgoing external edge appear. Number these new incoming external edges by $\alpha$ and $n+1$, and new outgoing external edge by $n+1$. New incoming external edge number $\alpha$ is included in the same paths as the old incoming external edge number $\alpha$. These paths pass through one new vertex. Then we must multiply by $1/2$. We get for the elements of the column number $\alpha$ of $\mathbf{a}(N+1)$  
\begin{equation}
\label{eq:18}
a_{r\alpha}(N+1)=(1/2)a_{r\alpha }(N) \textrm{,}
\end{equation}
where $\alpha$ is fixed, and $r$ ranges from 1 to $n$. New incoming external edge number $n+1$ is included in the same paths as new incoming external edge number $\alpha$. We get new column number $n+1$ with the following elements.
\begin{equation}
\label{eq:19}
a_{r (n+1)}(N+1)= a_{r \alpha}(N+1) \textrm{,}
\end{equation}
where $\alpha$ is fixed, and $r$ ranges from 1 to $n$. The new outgoing external edge number $n+1$ is connected by directed paths only with the incoming external edges numbers $\alpha$ and $n+1$. Each connection includes one path that passes through one vertex. We get new row number $n+1$ with the following elements.
\begin{equation}
\label{eq:20}
a_{(n+1) \alpha}(N+1)= a_{(n+1) (n+1)}(N+1)=1/2 \textrm{,}
\end{equation}
where $\alpha$ is fixed.
\begin{equation}
\label{eq:21}
a_{(n+1) \beta}(N+1)=0 \textrm{,}
\end{equation}
where $\beta$ ranges from $1$ to $\alpha-1$ and from $\alpha+1$ to $n$. The size of matrix of amplitudes is increased by 1 from $n$ to $n+1$.

We can calculate the probability of any elementary extension of any finite connected x-graph by finite number of steps of this algorithm. This algorithm is useful for numerical simulation. But it includes the matrixes of variable sizes. This is not useful for analytical investigations. Consider another form of this algorithm.

\section{Elementary evolution operators}
\label{EO}
Consider a finite sequential growth. The result of this growth is a finite x-graph $\mathcal{G}_N$ that includes $N$ vertices. Let $n$ be the number of outgoing or incoming external edges in $\mathcal{G}_N$. We can consider $\mathcal{G}_N$ as the result of $N$ steps of sequential growth from the empty x-graph.

Define modified matrixes $\mathbf{A}$ of amplitudes with the same size $n$. Let the matrix $\mathbf{a}(S)$ of amplitudes have a size $n(S)\leq n$ in the step number $S< N$. By definition, put $A_{i \alpha}(S)= a_{i \alpha}(S)$ if $i\leq n(S)$ and $\alpha \leq n(S)$, other diagonal elements of $\mathbf{A}(S)$ are equal to 1, other off-diagonal elements of $\mathbf{A}(S)$ are equal to 0. We have
\begin{equation}
\label{eq:22}
\begin{array}{c}
\mathbf{A}(S)=
\left( \begin{array}{cccc}
\begin{tabular}{|c|}\hline \\$\ \mathbf{a}(S)\ $\\ \\ \hline\end{tabular}&  &        & \\
&1&        & \\
&  &\dots& \\
&  &        &1
\end{array} \right) 
\textrm{.}
\end{array}
\end{equation}

Consider an elementary evolution operator. This is a following matrix.
\begin{equation}
\label{eq:23}
\begin{array}{c}
\mathbf{e}(ij)=\begin{array}{c}
 \\ \\ \\ i \\ \\ \\ \\ j \\ \\ \\ \\
\end{array}
\left( \begin{array}{ccccccccccc}
1&        &  &     &  &        &  &     &  &        & \\
  &\dots&  &     &  &        &  &     &  &         & \\
  &        &1&     &  &        &  &     &  &        & \\
  &        &  &1/2&  &        &  &1/2&  &        & \\
  &        &  &     &1&        &  &     &  &        & \\
  &        &  &     &  &\dots&  &     &  &        & \\
  &        &  &     &  &        &1&     &  &        & \\
  &        &  &1/2&  &        &  &1/2&  &        & \\
  &        &  &     &  &        &  &     &1&        & \\
  &        &  &     &  &        &  &     &  &\dots& \\
  &        &  &     &  &        &  &     &  &        &1
 \end{array} \right) 
\textrm{.}
\end{array}
\end{equation}
The elements ${e}_{ii}(ij)$, ${e}_{ij}(ij)$, ${e}_{ji}(ij)$, and ${e}_{jj}(ij)$ are equal to $1/2$. Other diagonal elements of $\mathbf{e}(ij)$ are equal to 1. Other off-diagonal elements of $\mathbf{e}(ij)$ are equal to 0.

If the step number $S+1$ is the addition of a new vertex to two outgoing external edges numbers $i$ and $j$, we have
\begin{equation}
\label{eq:24}
\mathbf{A}(S+1)= \mathbf{e}(ij) \mathbf{A}(S)\textrm{.}
\end{equation}

If the step number $S+1$ is the addition of a new vertex to one outgoing external edge number $i$, we have
\begin{equation}
\label{eq:25}
\mathbf{A}(S+1)= \mathbf{e}(i\ (n(S)+1)) \mathbf{A}(S)\textrm{.}
\end{equation}

If the step number $S+1$ is the addition of a new vertex to two incoming external edges numbers $\alpha$ and $\beta$, we have
\begin{equation}
\label{eq:26}
\mathbf{A}(S+1)= \mathbf{A}(S) \mathbf{e}(\alpha\beta)  \textrm{.}
\end{equation}

If the step number $S+1$ is the addition of a new vertex to one incoming external edge number $\alpha$, we have
\begin{equation}
\label{eq:27}
\mathbf{A}(S+1)= \mathbf{A}(S) \mathbf{e}(\alpha\ (n(S)+1)) \textrm{.}
\end{equation}

The matrix $\mathbf{A}(0)$ of the empty x-graph is a unity matrix $\mathbf{I}$ of size $n$. We have one vertex in the first step (Fig.\ \ref{fig:fig1}a). We get

\begin{equation}
\label{eq:28}
\mathbf{A}(1)= \mathbf{e}(1\ 2) \textrm{.}
\end{equation}

The evolution of the modified matrix of amplitudes is described as a sequence of the elementary evolution operators.
\begin{equation}
\label{eq:29}
\mathbf{A}(N)=\prod_{r=1}^N \mathbf{e}_r(i_r j_r)  \textrm{.}
\end{equation}

\section{Properties of the sequential growth}
\label{PSG}
Two elementary evolution operators $\mathbf{e}(ij)$ and $\mathbf{e}(bc)$ do not commute if $i=b$ and $j\ne c$, or $i=c$ and $j\ne b$. Otherwise they commute. If elementary evolution operators commute, we can add respective vertices in arbitrary order and get the same x-graph. In general case, otherwise is not truth. If elementary evolution operators do not commute, some respective vertices can be added in arbitrary order such that we get the same x-graph. Perhaps we get the different numbering of external edges.

\begin{law2}\label{T2}
The maximal value of an element of matrixes $\mathbf{p}_f$ and $\mathbf{p}_p$ is equal to $1/2$. 
\end{law2}

\textit{Proof.} The maximal value of an element of $\mathbf{a}$ is equal to $1/2$. The sum of the elements in each row and in each column of $\mathbf{a}$ is equal to 1. Any element of the matrix $\mathbf{p}_f$ is equal to the product of two rows of $\mathbf{a}$. Any element of the matrix $\mathbf{p}_p$ is equal to the product of two columns of $\mathbf{a}$. These products cannot be greater than the product of the maximal element of $\mathbf{a}$ and the sum of elements of row (of column) of $\mathbf{a}$. $\Box$

An element of the matrix $\mathbf{p}_f$ is equal to $1/2$ in one case. We get new outgoing external edge number $n+1$ by the elementary extension of the fourth type (Fig.\ \ref{fig:fig2}d). The probability $p_{(n+1)(n+1)}$ to add new vertex to this edge is equal to $1/2$. Similarly, an element of the matrix $\mathbf{p}_p$ is equal to $1/2$ in one case. We get new incoming external edge number $n+1$ by the elementary extension of the second type (Fig.\ \ref{fig:fig2}b). The probability $p_{(n+1)(n+1)}$ to add new vertex to this edge is equal to $1/2$ too.

\begin{figure}[t]
	\centering	
		\includegraphics[width=4cm,trim=8cm 17cm 8cm 7cm]{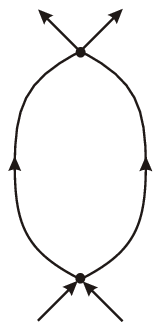}
	\caption{A double edge.}
	\label{fig:fig5}
\end{figure}

An observer cannot directly measure the structure of the x-graph. He can only calculate probabilities to get some structures in the series of identical experiments. If two different structures have the same matrixes of probabilities (\ref{eq:6}) - (\ref{eq:7}), an observer cannot distinguish them. This is the case if two different structures have the same matrixes of amplitudes. The simplest case is a double edge (Fig.\ \ref{fig:fig5}).
The matrix of amplitudes does not change if we add a double edge by the elementary extension of the first or third type. Respectively the elementary evolution operator is idempotent.
\begin{equation}
\label{eq:30}
\mathbf{e}(i\ j) \mathbf{e}(i\ j)= \mathbf{e}(i\ j) \textrm{.}
\end{equation}
The matrix of amplitudes does not change if we replace any vertex of the x-graph by two vertices that are connected by a double edge or if we do an inverse substitution. But we cannot exclude the generation of double edges from dynamics because this violate a normalization condition.

\section{Physical interpretations and perspectives}
\label{CON}
Consider the x-graph $\mathcal{G}_N$ with $n$ outgoing (incoming) external edges. Consider a sequential growth of this x-graph that only consists of elementary extensions of first and third types. In these elementary extensions we have the averaging of amplitudes (\ref{eq:9}) and (\ref{eq:17}). If elementary extensions can include every pairs of external edges, all amplitudes and probabilities (\ref{eq:2}) - (\ref{eq:3}) tend to $1/n$. If elementary extensions can include every pairs of external edges only in some subgraph, all amplitudes and probabilities of these elementary extensions tend to $1/n_1$, where $n_1$ is the number of outgoing (incoming) external edges in this subgraph. This result has clear physical meaning. Any closed system tends to thermodynamic equilibrium. All structures degrade.

Structures can emerge if there is an interaction with environment. The average probability is equal to $1/n$. In the case of big x-graph we have $1/n\ll 1$. The elementary extensions of second and fourth types generate elementary extensions with amplitudes and probabilities that equal to $1/2$ at the next step. The averaging with these amplitudes by elementary extensions of first and third types generates a set of elementary extensions with probabilities that much greater than other probabilities. These are preferable variants of the sequential growth. Probably such variants can generate self-organized structures. This is the task for further investigation.  

This model is useful for numerical simulation. There are first results \cite{FPP6}. We start from 1 vertex and calculate 500 steps. There are many variants of the growth for a big x-graph. But usually there are about very few variants with high probability. These are preferable variants of the growth. The maximal probability aperiodically oscillates during sequential growth. There are a variant with high probability in many steps. We hope that the existence of the small numbers of preferable variants of the growth is a symptom of self-organization. It is necessary to develop the methods to detect and analyze repetitive symmetrical self-organized structures during the numerical simulation of the sequential growth. This is the task for further investigation.

In the considered model, any physical processes are some structures of the x-graph. For physical interpretation we must determine the correspondence between physical quantities and properties of structures. Time is one of the most important physical quantities. In quantum theory time is measured by a macroscopic clock. This is the time of an observer. In the considered model, this is a sequence of addition of vertices. An observer can choose some structure as a clock. The number of vertices in the clock is a time interval by definition. For the measurement of time intervals of other processes we need procedure of comparison of instant of times (a synchronization of watches). In general case, processes include different numbers of vertices in the same time interval. Consequently we can describe any process by a frequency of discretization that is a ratio of the number of vertices in the process to a corresponding time interval. By definition, this frequency is equal to 1 for a clock. Similarly in quantum theory any particle is described by a frequency.

The transition from continuous spacetime to a causal set is a real quantization. We do not need any other quantization. We do not need a quantum dynamics of a causal set. Quantum properties must be consequences of the sequential growth. We have the evolution equation (\ref{eq:29}) for the matrix of amplitudes during sequential growth of an x-graph and the quadratic equations (\ref{eq:6}) - (\ref{eq:7}) for probabilities. This is like quantum theory. It is important that a causal set is a dyadic x-graph. Otherwise we cannot get the quadratic dependence of probabilities on amplitudes. But in this model, all numbers are real. It may be we can get complex amplitudes by Fourier transform of the considered amplitudes.

We can generalize this dynamics. We can consider the nonequal probabilities on the first step of the algorithm. This is the time asymmetry. We can consider the nonequal probabilities on the second step of the algorithm. This is the preferable growth of some subgraphs.

I am grateful to Alexander V. Kaganov and Vladimir V. Kassandrov for extensive discussions on this subject, and Ivan V. Stepanian for collaboration in a numerical simulation.

\end{document}